\documentclass[twocolumn,journal]{IEEEtran}

\newif\ifnocomment
\nocommenttrue

\usepackage{fixltx2e}
\usepackage{mathrsfs}
\usepackage{amsmath}
\usepackage{amsthm}
\usepackage{amssymb}
\usepackage{mathtools}
\usepackage{graphics}
\usepackage{float}
\usepackage{epstopdf}
\usepackage{graphicx}
\usepackage{bm}
\usepackage{tikz}
\usepackage[lined,boxed,commentsnumbered,linesnumbered,ruled]{algorithm2e}
\usepackage{cite}
\usepackage{comment}
\usepackage{balance}
\usepackage{booktabs} 
\usepackage{enumitem}
\usepackage{mathtools}
\usetikzlibrary{shapes.geometric,shapes, shapes.misc, patterns, decorations.text,plotmarks,3d, positioning,calc,arrows.meta,intersections}
\usepackage{pgfplots}
\usepgfplotslibrary{fillbetween} 
\usepackage{ifthen}
\usepackage{multirow}
\usepackage{hyperref} 
\usepackage[capitalize]{cleveref}
\crefname{equation}{\unskip}{\unskip}
\usepackage{acronym}

\newtheorem{theorem}{Theorem}
\newtheorem{definition}{Definition}
\newtheorem{lemma}{Lemma}

\newtheorem{proposition}{Proposition}

\newcommand{\kmax}{k_\mathsf{max}}
\newcommand{\dmax}{\delta_\mathsf{max}}
\newcommand{\kmin}{k_\mathsf{min}}
\newcommand{\mumin}{\mu_\mathsf{min}}
\newcommand{\mumax}{\mu_\mathsf{max}}
\newcommand{\const}[1]{\textnormal{\usefont{U}{eur}{m}{n}\selectfont #1}}

\newcommand{\GF}{\mathrm{GF}}

\newcommand{\NSBS}{N_\mathsf{SBS}}
\newcommand{\bp}{\boldsymbol{p}}
\newcommand{\bX}{\boldsymbol{X}}
\newcommand{\bXi}[1]{\boldsymbol{X}^{(#1)}}
\newcommand{\bCi}[1]{\boldsymbol{C}^{(#1)}}
\newcommand{\bxij}[2]{\boldsymbol{x}^{(#1)}_{#2}}

\newcommand{\btxij}[2]{\tilde{\boldsymbol{x}}^{(#1)}_{#2}}

\newcommand{\bcij}[2]{\boldsymbol{c}^{(#1)}_{#2}}

\newcommand{\bcijj}[3]{c^{(#1)}_{#2,#3}}
\newcommand{\brl}[1]{\boldsymbol{r}^{(#1)}}
\newcommand{\bql}[1]{\boldsymbol{q}^{(#1)}}
\newcommand{\rl}[1]{r^{(#1)}}

\newcommand{\RPIR}{\const{R}_\mathsf{PIR}}
\newcommand{\RPIRpop}{\const{R}_\mathsf{PIR}^\mathsf{pop}}
\newcommand{\RnPIRpop}{\const{R}_\mathsf{noPIR}^\mathsf{pop}}
\newcommand{\RnPIR}{\const{R}_\mathsf{noPIR}}

\newcommand{\DPIR}{\const{D}_\mathsf{PIR}}
\newcommand{\CPIR}{\const{C}_\mathsf{PIR}}

\newcommand{\C}{\mathcal{C}}
\newcommand{\Ctilde}{\tilde{\mathcal{C}}}

\newcommand{\CMDS}{\mathcal{C}_{\mathsf{MDS}}^{\bm \mu}}
\newcommand{\Cmax}{\mathcal{C}_\mathsf{max}}

\newcommand{\Hwt}[1]{w_\mathsf{H}\left(#1\right)}   

\acrodef{MBS}{macro base station}
\acrodef{BS}{base station}
\acrodef{SBS}{small-cell base station}
\acrodefplural{SBS}[SBSs]{small-cell base stations}
\acrodef{MDS}{maximum distance separable}
\acrodef{PIR}{private information retrieval}
\acrodef{DSS}{Distributed Storage System}
\acrodef{ML}{maximum likelihood}
\acrodef{BEC}{binary erasure channel}

\makeatletter
\tikzoption{canvas is xy plane at z}[]{%
  \def\tikz@plane@origin{\pgfpointxyz{0}{0}{#1}}%
  \def\tikz@plane@x{\pgfpointxyz{1}{0}{#1}}%
  \def\tikz@plane@y{\pgfpointxyz{0}{1}{#1}}%
  \tikz@canvas@is@plane
}
\makeatother

\makeatletter
\newcommand*\rel@kern[1]{\kern#1\dimexpr\macc@kerna}
\newcommand*\widebar[1]{%
  \begingroup
  \def\mathaccent##1##2{%
    \rel@kern{0.8}%
    \overline{\rel@kern{-0.8}\macc@nucleus\rel@kern{0.2}}%
    \rel@kern{-0.2}%
  }%
  \macc@depth\@ne
  \let\math@bgroup\@empty \let\math@egroup\macc@set@skewchar
  \mathsurround\z@ \frozen@everymath{\mathgroup\macc@group\relax}%
  \macc@set@skewchar\relax
  \let\mathaccentV\macc@nested@a
  \macc@nested@a\relax111{#1}%
  \endgroup
}
\makeatother

\usetikzlibrary{external}
\begin{document}

\title{Private Information Retrieval From a Cellular Network With Caching at the Edge} %

\author{Siddhartha Kumar,~\IEEEmembership{Student Member,~IEEE},  Alexandre Graell i Amat,~\IEEEmembership{Senior Member,~IEEE}, \\Eirik Rosnes,~\IEEEmembership{Senior Member,~IEEE}, and Linda Senigagliesi,~\IEEEmembership{Student Member,~IEEE}
\thanks{S. Kumar and E.\ Rosnes were supported by the Research Council of Norway (grant 240985/F20). A.\ Graell i Amat was supported by the Swedish Research Council under grant \#2016-04253.}
\thanks{S. Kumar and E. Rosnes are with Simula UiB, N-5020 Bergen, Norway (e-mail: \{kumarsi,eirikrosnes\}@simula.no).}
\thanks{A. Graell i Amat is with the Department of Electrical Engineering, Chalmers University of Technology, SE-41296 Gothenburg, Sweden (e-mail: alexandre.graell@chalmers.se).}
\thanks{L. Senigagliesi is with Dipartimento di Ingegneria dell'Informazione, Universit\`a Politecnica delle Marche, Ancona, Italy (e-mail: l.senigagliesi@pm.univpm.it).}

}


\maketitle

\begin{abstract}


We consider the problem of downloading content from a cellular network where content is cached at the wireless edge while achieving privacy.
In particular, we consider \ac{PIR} of content from a library of files, i.e.,  the user wishes to download a file and does not want the network to learn any information about which file she is interested in. To reduce the backhaul usage, content is cached at the wireless edge in a number of \acp{SBS} using maximum distance separable codes. We propose a PIR scheme for this scenario that achieves privacy against a number of spy \acp{SBS} that (possibly) collaborate. The proposed PIR scheme is an extension of a recently introduced scheme by Kumar \emph{et al.} to the case of multiple code rates, suitable for the scenario where files have different popularities. We then derive the backhaul rate and optimize the content placement to minimize it. We prove that uniform content placement is optimal, i.e., all files that are cached should be stored using the same code rate. This is in contrast to the case where no PIR is required. Furthermore, we show numerically that popular content placement is optimal for some scenarios. 
\end{abstract}

\section{Introduction}

Bringing content closer to the end user in wireless networks, the so-called caching at the wireless edge, has emerged as a promising technique to reduce the backhaul usage. The literature on wireless caching is vast. Information-theoretic aspects of caching were studied in \cite{Nie12,Mad14}. To leverage the potential gains of caching, several papers proposed to cache files in densely deployed small-cell base stations (SBSs) with large storage capacity, see, e.g., \cite{And14,Liu16,Sha13,Bas14,Bio15}. In \cite{Sha13}, content is cached in SBSs using maximum distance separable (MDS) codes to reduce the download delay. This scenario was further studied in \cite{Bio15}, where the authors optimized the MDS-coded caching to minimize the backhaul rate. Caching content directly in the mobile devices and exploiting device-to-device communication has been considered in, e.g., \cite{Ji16, Gol14, Ped16,Pie16,Ped18}.

Recently, private information retrieval (PIR) has attracted a significant interest in the research community \cite{Ish04,Sha14,Cha15,Taj16,Kum17b,Sun17,Sun17b,Hol17b,Kum18,Ban18,Sun18}. In PIR, a user would like to retrieve data from a distributed storage system (DSS) in the presence of spy nodes, without revealing any information about the piece of data she is interested in to the spy nodes. PIR was first studied by Chor \emph{et al.} \cite{cho95} for the case where a binary database is replicated among $n$ servers (nodes) and the aim is to privately retrieve a single bit from the database in the presence of a single spy node (referred to as the noncolluding case), while minimizing the total communication cost. In the last few years, spurred by the rise of DSSs, research on PIR has been focusing on the more general case where data is stored using a storage code. 

The PIR capacity, i.e., the maximum achievable PIR rate, was studied in \cite{Sun17,Sun17b,Kum18,Ban18,Sun18}. In \cite{Sun17b,Sun18}, the PIR capacity was derived for the scenario where data is stored in a DSS using a repetition code. In \cite{Ban18}, for the noncolluding case, the authors derived the PIR capacity for the scenario where data is stored using an (single) MDS code, referred to as the MDS-PIR capacity. For the case where several spy nodes collaborate with each other, referred to as the colluding case, the MDS-PIR capacity is in general still unknown, except for some special cases \cite{Sun17} (and for repetition codes  \cite{Sun18}). 
PIR protocols for DSSs have been proposed in \cite{Sha14,Taj16,Kum17b,Hol17b,Kum18}. 
In \cite{Taj16}, a PIR protocol for MDS-coded DSSs was proposed and shown to achieve the MDS-PIR capacity for the case of noncolluding nodes when the number of files stored in the DSS goes to infinity. PIR protocols for the case where data is stored using non-MDS codes were proposed in \cite{Kum17b,Hol17b,Kum18}. 

In this paper, we consider PIR of content from a cellular network. In particular, we consider the private retrieval of content from a library of files that have different popularities. We consider a similar scenario as in \cite{Bio15} where, to reduce the backhaul usage, content is cached in SBSs using MDS codes. We propose a PIR scheme for this scenario that achieves privacy against a number of spy \acp{SBS} that possibly collude. The proposed PIR scheme is an extension of Protocol~3 in \cite{Kum18} to the case of multiple code rates, suitable for the scenario where files have different popularities. We also propose an MDS-coded content placement slightly different than the one in \cite{Bio15} but that is more adapted to the PIR case. We show that, for the conventional content retrieval scenario with no privacy, the proposed content placement is equivalent to the one in \cite{Bio15}, in the sense that it yields the same average backhaul rate. We then derive the backhaul rate for the PIR case as a function of the content placement. We prove that uniform content placement, i.e., all files that are cached are encoded with the same code rate, is optimal. This is a somewhat surprising result, in contrast to the case where no PIR is considered, where optimal content placement is far from uniform \cite{Bio15}. We further consider the minimization of a weighted sum of the backhaul rate and the communication rate from the SBSs, relevant for the case where limiting the communication from the SBSs is also important. 
We finally report numerical results for both the scenario where SBSs are placed regularly in a grid and for a Poisson point process (PPP) deployment  model where SBSs are distributed over the plane according to a PPP.  We show numerically that popular content placement is optimal for some system parameters. To the best of our knowledge, PIR for the wireless caching scenario has not been considered before.
 
Notation: We use lower case bold letters to denote vectors, upper case bold letters
to denote matrices, and calligraphic upper case letters to denote sets. For example, $\bm x$, $\bm X$, and $\mathcal X$
denote a vector, a matrix, and a set, respectively.  We denote a submatrix of $\bm X$ that is restricted in columns by the set $\mathcal{I}$ by $\bm{X}|_{\mathcal{I}}$. 
$\C$ will denote a linear code over the finite field $\GF(q)$. The multiplicative subgroup of $\GF(q)$ (not containing the zero element) is denoted by $\GF(q)^{\times}$. 
We use the customary code parameters $(n,k)$ to denote a code $\C$ of blocklength $n$ and dimension $k$. A generator matrix for $\C$ will be denoted by $\bm G^{\C}$ and a parity-check matrix by  $\bm H^{\C}$. A set of coordinates of $\C$, $\mathcal{I}\subseteq \{1,\ldots,n\}$, of size $k$ is said to be an \emph{information set} if and
only if $\bm{G}^{\C}|_\mathcal{I}$ is invertible. 
%
%
The Hadamard product of two linear subspaces $\C$ and $\C'$, denoted by $\C\circ\C'$, is the space generated by the Hadamard products $\bm c \circ \bm c' \triangleq (c_1c_1',\ldots,c_nc_n')$ for all pairs $\bm c \in \C$, $\bm c' \in \C'$. The inner product of two vectors $\bm x$ and $\bm x'$ is denoted by $\langle \bm x,\bm x' \rangle$, while $\Hwt{\bm{x}}$ denotes the Hamming weight of $\bm{x}$. 
$(\cdot)^\top$ represents the transpose of its argument, while $\mathsf H(\cdot)$ represents the entropy function. 
%
With some abuse of language, we sometimes interchangeably refer to binary vectors as erasure patterns under the implicit assumption that the ones represent erasures. An erasure pattern  (or binary vector) $\bm{x}$ is said to be correctable by a code $\C$ if matrix $\bm H^{\C}|_{\chi(\bm{x})}$ has rank $|\chi(\bm{x})|$.

\section{System Model}
\label{sec:SystemModel}

\begin{figure}
\centering
\includegraphics[width=1\columnwidth]{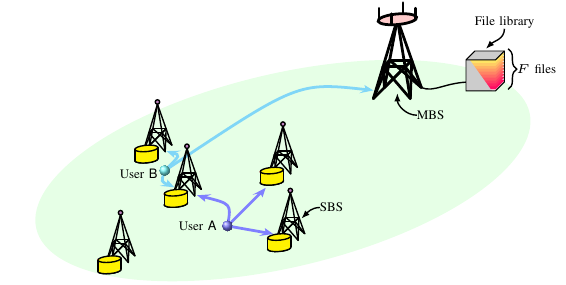}
\caption{A wireless network for content delivery consisting of a MBS and five SBSs. Users download files from a library of $F$ files. The MBS has access to the library through a backhaul link. Some files are also cached at SBSs using a $(5,3)$ MDS code. User \textsf A retrieves a cached file from the three SBSs within range. User \textsf B retrieves a fraction $2/3$ of a cached file from the two SBSs within range and the remaining fraction from the MBS.}
\label{Fig:SystemModel}
\end{figure}

We consider a cellular network where a macro-cell is served by a macro base station (MBS). Mobile users wish to download files from a library of $F$ files that is always available at the MBS through a backhaul link. We assume all files of equal size.\footnote{Assuming files of equal size is without loss of generality, since content can always be divided into chunks of equal size.} In particular, each file consists of $\beta L$ bits and is represented by a $\beta\times L$ matrix $\bXi{i}$, 
\begin{align*}
\bX{^{(i)}}=\left(\begin{array}{c} \btxij{i}{1}  \\ \vdots \\ \btxij{i}{\beta} \end{array}\right)
\end{align*}
where upperindex $i=1,\ldots,F$ is the file index. Therefore, each file can be seen as divided into $\beta$ stripes $\btxij{i}{1},\ldots,\btxij{i}{\beta}$ of $L$ bits each. The file library has popularity distribution $\bp=(p_1,\ldots,p_F)$, where file $\bX^{(i)}$ is requested with probability $p_i$. We also assume that $\NSBS$ \acp{SBS} are deployed to serve requests and offload traffic from the MBS whenever possible. To this purpose, each \ac{SBS} has a cache size equivalent to $M$ files. The considered scenario is depicted in Fig.~\ref{Fig:SystemModel}.

\subsection{Content Placement} 

File $\bXi{i}$ is partitioned into $\beta k_i$ packets of size $L/k_i$ bits and encoded before being cached in the \acp{SBS}. In particular, each packet is mapped onto a symbol of the  field $\GF(q^{\delta_i})$, with $\delta_i\ge  \frac{L}{k_i\log_2 q }$. For simplicity, we assume that $\frac{L}{k_i\log_2 q }$ is integer and set $\delta_i=  \frac{L}{k_i\log_2 q }$. Thus, stripe $\btxij{i}{a}$ can be equivalently represented by a  stripe $\bxij{i}{a}$, $a=1,\ldots,\beta$, of symbols  over $\GF(q^{\delta_i})$. Each stripe $\bxij{i}{a}$ is then encoded using an $(\NSBS,k_i)$ MDS code $\C_i$ over $\GF(q)$ into a codeword $\bcij{i}{a}=(\bcijj{i}{a}{1},\ldots,\bcijj{i}{a}{\NSBS})$, where code symbols $\bcijj{i}{a}{j}$, $j=1,\ldots,\NSBS$, are over $\GF(q^{\delta_i})$. For later use, we define $\kmin\triangleq\min\{k_i\}$, $\kmax\triangleq\max\{k_i\}$, and  $\dmax\triangleq\frac{L}{\kmin\log_2 q }$. 

The encoded file can be represented by a $\beta\times \NSBS$ matrix $\bCi{i}=(c^{(i)}_{a,j})$. Code symbols $c^{(i)}_{a,j}$ are then stored in the $j$-th SBS (the ordering is unimportant). Thus, for each file $\bXi{i}$, each \ac{SBS} caches one coded symbol of each stripe of the file, i.e., a fraction $\mu_i=1/k_i$ of the $i$-th file. As $k_i \in \{1,\ldots,\NSBS-1\}$, 
\begin{align*}
\mu_i\in\mathcal{M}\triangleq\{0,1/(\NSBS-1),\ldots,1/2,1\},
\end{align*} 
where $\mu_i=0$ implies that file $\bXi{i}$ is not cached. Note that, to achieve privacy, $k_i<\NSBS$, i.e., files need to be cached with redundancy. As a result, $\mu_i=1/\NSBS$ is not allowed. This is in contrast to the case of no PIR, where $k_i=\NSBS$ (and hence $\mu_i=1/\NSBS$) is possible.

Since each \ac{SBS} can cache the equivalent of $M$ files, the $\mu_i$'s must satisfy
\begin{align*}
	\sum_{i=1}^F \mu_i &\le  M .
\end{align*}

We define the vector $\bm \mu=(\mu_1,\ldots,\mu_F)$ and refer to it as the \emph{content placement}. Also, we denote by $\CMDS$ the caching scheme that uses MDS codes $\{\C_i\}$ according to the content  placement $\bm \mu$. For later use, we define $\mumin\triangleq \min\{\mu_i|\mu_i\neq 0\}$ and $\mumax\triangleq \max\{\mu_i\}$.

We remark that the content placement above is slightly different than the content placement proposed in \cite{Bio15}. In particular, we assume fixed code length (equal to the number of \acp{SBS}, $\NSBS$) and variable $k_i$, such that, for each file cached, each \ac{SBS} caches a single symbol from each stripe of the file. In \cite{Bio15}, the content placement is done by first dividing each file into $k$ symbols and encoding them using an $(\tilde{n}_i,k)$ MDS code, where $\tilde{n}_i=k+(N_{\mathsf{SBS}}-1)m_i$, $m_i\le k$. Then, $m_i$ (different) symbols of the $i$-th file are stored in each \ac{SBS} and the MBS stores $k-m_i$ symbols.\footnote{This is because the model in \cite{Bio15} assumes that one \ac{SBS} is always accessible to the user. If this is not the case, the MBS must store all $k$ symbols of the file. Here, we consider the case where the MBS must store all $k$ symbols because it is a bit more general.} Our formulation is perhaps a bit simpler and more natural from a coding perspective. Furthermore, we will show in Section~\ref{sec:CAnPIR} that the proposed content placement is equivalent to the one in \cite{Bio15}, in the sense that it yields the same average backhaul rate.

\subsection{File Request}

Mobile devices request files according to the popularity distribution $\bp=(p_1,\ldots,p_F)$. Without loss of generality, we assume $p_1\ge p_2\ge \ldots \ge p_F$. The user request is initially served by the \acp{SBS} within communication range. We denote by $\gamma_b$ the probability that the user is served by $b$ \acp{SBS} and define $\bm\gamma=(\gamma_0,\ldots,\gamma_{\NSBS})$. If the user is not able to completely retrieve $\bXi{i}$ from the \acp{SBS}, the additional required symbols are fetched from the MBS. Using the terminology in \cite{Bio15}, the average fraction of files that are downloaded from the MBS is referred to as the backhaul rate, denoted by $\const{R}$, and defined as
\begin{align*}
\const{R}\triangleq\frac{\text{average no. of bits downloaded from the MBS}}{\beta L}.
\end{align*}
Note that for the case of no caching $\const{R}=1$.

As in \cite{Bio15}, we assume that the communication is error free.

\subsection{Private Information Retrieval and Problem Formulation}
We assume that some of the \acp{SBS} are spy nodes that (potentially) collaborate with each other. On the other hand, we assume that the MBS can be trusted. The users wish to retrieve files from the cellular network, but do not want the spy nodes to learn any information about which file is requested by the user. The goal is to retrieve data from the network privately while minimizing the use of the backhaul link, i.e., while minimizing $\const{R}$. Thus, the goal is to optimize the content placement $\bm \mu$ to minimize $\const{R}$.
 
\section{Private Information Retrieval Protocol}
\label{sec:PIRprotocol}

In this section, we present a PIR protocol for the caching scenario. The PIR protocol proposed here is an extension of Protocol 3 in \cite{Kum18} to the case of multiple code rates.\footnote{Protocol 3 in \cite{Kum18} is based on and improves the protocol in \cite{Hol17b}, in the sense that it achieves higher PIR rates.}

Assume without loss of generality that the user wants to download file $\bXi{i}$. To retrieve the file, the user generates $n\le \NSBS$ query matrices, $\bm Q^{(l)}$, $l=1,\ldots, n$, where $\bm Q^{(1)},\ldots, \bm Q^{(b)}$ are the queries sent to the $b$ SBSs within visibility and the remaining $n-b$ queries $\bm Q^{(b+1)},\ldots, \bm Q^{(n)}$ are sent to the MBS. Note that $n$ is a parameter that needs to be optimized. Each query matrix is of size $d\times \beta F$ symbols (from $\GF(q)$) and has the following structure,
\begin{align*}
	\bm Q^{(l)}=\left(\begin{matrix}
		\bm q^{(l)}_1\\
		\bm q^{(l)}_2\\
		\vdots\\
		\bm q^{(l)}_d\\
	\end{matrix}\right)=\left(\begin{matrix}
		q^{(l)}_{1,1} & q^{(l)}_{1,2} & \cdots & q^{(l)}_{1,\beta F}\\
		q^{(l)}_{2,1} & q^{(l)}_{2,2} & \cdots & q^{(l)}_{2,\beta F}\\
		\vdots & \vdots & \cdots & \vdots\\
		q^{(l)}_{d,1} & q^{(l)}_{d,2} & \cdots & q^{(l)}_{d,\beta F}\\
	\end{matrix}\right).
\end{align*}
The query matrix $\bm Q^{(l)}$ consists of $d$ subqueries $\bm q^{(l)}_j$, $j=1,\ldots,d$,
of length $\beta F$ symbols each. In response to query matrix $\bm Q^{(l)}$, a SBS (or the MBS) sends back to the user a response vector $\brl{l}=(\rl{l}_1,\ldots, \rl{l}_d)^\top$  of length $d$, computed as
\begin{align}
\label{eq:response}
	\brl{l}=(\rl{l}_1,\ldots, \rl{l}_d)^\top = \bm
  Q^{(l)}\bigl(c^{(1)}_{1,l},\ldots,c^{(1)}_{\beta,l},\ldots,c^{(F)}_{\beta,l}\bigr)^\top.
\end{align}
We will denote the $j$-th entry of the response vector $\brl{l}$, i.e., $\rl{l}_j$, as the $j$-th subresponse of $\brl{l}$. Each response vector consists of $d$ subresponses, each being a linear combination of $\beta F$ symbols. Note that the operations are performed over the largest extension field, i.e., $\GF(q^{\dmax})$, and the subresponses are also over this field, i.e., each subresponse is of size $L/\kmin=L\mumax$ bits and hence each response is of size $dL\mumax$ bits.

The queries and the responses must be such that privacy is ensured and the user is able to recover the requested file. More precisely, information-theoretic PIR in the context of wireless caching with spy SBSs is defined as follows.
\begin{definition}
Consider a wireless caching scenario with $\NSBS$ SBSs that cache parts of a library of $F$ files and in which a set $\mathcal T$ of $T$ SBSs act as colluding  spies. A user wishes to
retrieve the $i$-th file and generates queries $\bm Q^{(l)}$, $l=1,\ldots,n$. In response to the queries the SBSs and (potentially) the MBS send back the 
responses $\bm r^{(l)}$. This scheme achieves perfect information-theoretic PIR if and only if
\begin{subequations} 
  \begin{align}
    \label{Def: cond1}
    \text{Privacy:}\qquad  &\mathsf H\bigl(i|\bm Q^{(l)},l\in\mathcal T\bigr)=\mathsf H(i);
    \\
    \label{Def: cond2}
    \text{Recovery:}\qquad &\mathsf H\bigl(\bm X^{(i)}|\bm r^{(1)},\ldots, \bm r^{(n)}\bigr)=0.
  \end{align}
\end{subequations}
\end{definition}
Condition \eqref{Def: cond1} means that the spy SBSs gain no additional information about which file is requested from the queries (i.e., the uncertainty about the file requested after observing the queries is identical to the a priori uncertainty determined by the popularity distribution), while Condition \eqref{Def: cond2} guarantees that the user is able to recover the file from the $n$ response vectors.

We define the $(n,k_i)$ code $\C'_i$, $i=1,\ldots,F$, as the code obtained by puncturing the underlying $(\NSBS,k_i)$ storage code $\C_i$, and by  $\Cmax'$ the code with parameters $(n,\kmax)$.\footnote{Without loss of generality, to simplify notation we assume that the last coordinates of the code are puntured.}  For the protocol to work, we require that $\kmin$ divides $k_i$ for all $i$, i.e., $\kmin\mid k_i$. This ensures that $\GF(q^{\delta_i})\subseteq\GF(q^{\dmax})$. Furthermore, we require the codes $\C'_i$ to be such that $\C'_i\subseteq \Cmax'$. The protocol is characterized by the codes $\{\C'_i\}$ and by two other codes, $\bar{\mathcal C}$ and $\tilde{\mathcal C}$. Code $\bar{\mathcal C}$ (over $\GF(q)$) has parameters $(n,\bar k)$ and characterizes the queries sent to the \acp{SBS} and the MBS, while code $\tilde{\mathcal C}$ (defined below) defines the responses sent back to the user from the \acp{SBS} and the MBS. The designed protocol achieves PIR against a number of colluding SBSs $T\le d_{\mathsf{min}}^{\bar{\mathcal C}^\perp}-1$, where $d_{\mathsf{min}}^{\bar{\mathcal C}^\perp}$ is the minimum Hamming distance of the dual code of $\bar{\mathcal C}$.

\subsection{Query Construction}

The queries must be constructed such that privacy is preserved and the user can retrieve the requested file from the $n$ response vectors $\brl{l}$, $l=1,\ldots,n$.
In particular, the protocol is designed such that the subresponses $\rl{l}_j$, $l=1,\ldots,n$, corresponding to the $n$ subqueries $\bm{q}_j^{(1)},\ldots,\bm{q}_j^{(n)}$ recover $\Gamma$ unique code symbols of the file $\bXi{i}$.

The queries are constructed as follows.
The user chooses  $\beta F$ codewords $\bar{\bm c}_m^{(i)}=(c_{m,1}^{(i)},\ldots,c_{m,n}^{(i)})\in\bar{\mathcal C}$, $m=1,\ldots,\beta$, $i=1,\ldots,F$, independently and uniformly at random. Then, the user constructs $n$ vectors,
\begin{align}
\label{eq:vector_circle}
  \mathring{\bm c}_{l}=(\mathring{\bm c}^{(1)}_l,\ldots,\mathring{\bm c}^{(F)}_l),\quad l=1,\ldots,n,
\end{align}
where $\mathring{\bm c}^{(i)}_l$ collects the $l$-th coordinates of  the $\beta$ codewords  $\bar{\bm c}_m^{(i)}$, $m=1,\ldots,\beta$, i.e., $\mathring{\bm c}^{(i)}_l=(\bar c^{(i)}_{1,l},\ldots,\bar c^{(i)}_{\beta,l})$.

Assume that the user wants to retrieve file $\bXi{i}$. Then, subquery $\bql{l}_j$ is constructed as
 \begin{align}
  \label{Eq: CollQuery_designa}
 \bm q_j^{(l)}=\mathring{\bm{c}}_{l}+\bm \delta_j^{(l)}, 
 \end{align}
 where
   \begin{align}
    \label{Eq: CollQuery_design}
    \bm \delta_j^{(l)}=\begin{cases}
      \bm \omega_{\beta(i-1)+s_j^{(l)}} & \text{if }
      l\in\mathcal J_{j},
      \\
     \bm \omega_{0} & \text{otherwise},
    \end{cases}
  \end{align}
  for some set $\mathcal J_{j}$ that will be defined below. Vector $\bm \omega_t$, $t=1,\ldots,\beta F$, denotes the $t$-th $(\beta F)$-dimensional unit vector, i.e., the length-$\beta F$ vector with a one in the $t$-th coordinate and zeroes in all other coordinates, and $\bm \omega_0$ the all-zero vector. The meaning of index $s_{j}^{(l)}$ will become apparent later. 


According to \eqref{Eq: CollQuery_designa}, each subquery vector is the sum of two vectors, $\mathring{\bm{c}}_{l}$ and $\bm \delta_j^{(l)}$. The purpose of $\mathring{\bm{c}}_{l}$ is to make the subquery appear random and thus ensure privacy (i.e., Condition~\eqref{Def: cond1}). On the other hand, the vectors $\bm \delta_j^{(l)}$ are deterministic vectors which must be properly constructed such that the user is able to retrieve the requested file from the response vectors (i.e., Condition~\eqref{Def: cond2}). Similar to Protocol 3 in \cite{Kum18}, the vectors $\bm \delta_j^{(l)}$ are constructed from a $d\times n$ binary matrix $\hat{\bm E}$ where each row represents a weight-$\Gamma$ erasure pattern 
that is correctable by $\Ctilde$ and where the weights of its columns are determined from
$\beta$ information sets $\mathcal{I}_m$, $m=1,\ldots,\beta$, of $\Cmax'$.

The construction of $\hat{\bm E}$ is addressed below.  
We define the set $\mathcal F_l$ as the index set of information sets $\mathcal I_m$ that contain the $l$-th coordinate of $\Cmax'$, i.e., $\mathcal F_l=\{m:l\in\mathcal I_m\}$.
To allow the user to recover the requested file from the response vectors,  $\hat{\bm E}$ is constructed such that it satisfies the following conditions.
\begin{enumerate}
\item[$\mathsf{C1.}$] The user should be able to recover $\Gamma$ unique code symbols of the requested file
  $\bm{X}^{(i)}$ from the responses to each set of $n$ subqueries $\bm q^{(l)}_j$, $l=1,\ldots,n$. This is to say that each row of $\hat{\bm E}$ should have exactly $\Gamma$ ones. We denote by $\mathcal J_j$  the support of the $j$-th row of $\hat{\bm E}$.
  \item[$\mathsf{C2.}$] The user should be able to recover  $\Gamma d\ge \beta k_i$ unique code symbols of the requested file
  $\bm X^{(i)}$, at least $k_i$ symbols from each stripe. This means that each row $\hat{\bm{e}}_j=(\hat e_{j,1},\ldots,\hat e_{j,n})$, $j=1,\ldots,d$, of $\hat{\bm{E}}$    should correspond to an erasure pattern 
  that is correctable by $\tilde{\mathcal C}$.
\item[$\mathsf{C3.}$] Let $\bm{t}_l$, $l=1,\ldots,n$, be the $l$-th column vector of $\hat{\bm{E}}$. 
   The protocol should be able to recover $\Hwt{\bm{t}_l}$ unique
  code symbols from the $l$-th response vector, which means that it is required that $\Hwt{\bm{t}_l}= |{\mathcal{F}_l}|$. We call the vector $(\Hwt{\bm{t}_1},\ldots,\Hwt{\bm{t}_n})$ the \emph{column weight
    profile of $\hat{\bm{E}}$}.
\end{enumerate}

Finally, from $\hat{\bm{E}}$ we construct the vectors $\bm \delta_j^{(l)}$ 
in \eqref{Eq: CollQuery_design}. In particular, index  $s_{j}^{(l)}$ in \eqref{Eq: CollQuery_design} is such that $s_j^{(l)}\in\mathcal F_l$ and $s_{j}^{(l)} \neq s_{j'}^{(l)}$ for $j \neq j'$, $j,j' = 1,\ldots,d$.

\subsection{Response Vectors}

The $j$-th subresponse corresponding to subquery $\bql{l}_j$, $j=1,\ldots, d$, is (see \eqref{eq:response})
\begin{align*}
	\rl{l}_j=\langle \bm q_j^{(l)}, (c_{1,l}^{(1)},\ldots,c_{\beta,l}^{(F)})\rangle.
\end{align*}

The user collects the $n$ subresponses $\rl{l}_j$, $l=1,\ldots n$, in the vector $\bm\rho_j$,
\begin{align}
\label{Eq: PIR_coll_response}
  \bm \rho_{j}=\left(\begin{matrix}
      \rl{1}_j\\
      \rl{2}_j\\
      \vdots\\
      \rl{n}_j
    \end{matrix}\right)=&
  \sum_{m=1}^\beta\underbrace{\left(\begin{matrix}
      \bar c_{m,1}^{(1)}c_{m,1}^{(1)}\\
      \bar c_{m,2}^{(1)}c_{m,2}^{(1)}\\
      \vdots\\
      \bar c_{m,n}^{(1)}c_{m,n}^{(1)}\\
    \end{matrix}\right)}_{\substack{\in\, \left\{  \bm x \in  (\GF(q^{\delta_{\rm max}}))^n \colon \right. \\ \left. \bm H^{\C'_1\circ\,\bar{\mathcal C}} \bm x = \bm 0  \right\}  }}  
    +\underbrace{
  \left(\begin{matrix}
      \bar c_{m,1}^{(2)}c_{m,1}^{(2)}\\
      \bar c_{m,2}^{(2)}c_{m,2}^{(2)}\\
      \vdots\\
      \bar c_{m,n}^{(2)}c_{m,n}^{(2)}\\
    \end{matrix}\right)}_{\substack{\in\, \left\{  \bm x \in  (\GF(q^{\delta_{\rm max}}))^n \colon \right. \\ \left. \bm H^{\C'_2\circ\,\bar{\mathcal C}} \bm x = \bm 0  \right\}  }} \nonumber \\
    &+\cdots+
   \underbrace{
  \left(\begin{matrix}
      \bar c_{m,1}^{(F)}c_{m,1}^{(F)}\\
      \bar c_{m,2}^{(F)}c_{m,2}^{(F)}\\
      \vdots\\
      \bar c_{m,n}^{(F)}c_{m,n}^{(F)}\\
    \end{matrix}\right)}_{\substack{\in\, \left\{  \bm x \in  (\GF(q^{\delta_{\rm max}}))^n \colon \right. \\ \left. \bm H^{\Cmax' \circ\,\bar{\mathcal C}} \bm x = \bm 0  \right\}  }}
  +\left(\begin{matrix}
      o_j^{(1)}\\
      o_j^{(2)}\\
      \vdots\\
      o_j^{(n)}
  \end{matrix}\right),
\end{align}
where symbol $o^{(l)}_j$ represents the code symbol from file $\bXi{i}$ downloaded in the $j$-th subresponse from the $l$-th response vector. Due to the structure of the queries obtained from $\hat{\bm E}$, the user retrieves $\Gamma$ code symbols from the set of $n$ subresponses to the $j$-th subqueries. Consider a retrieval code $\tilde{\mathcal C}$ of the form
\begin{align}
\label{Eq: Ctilde}
	\tilde{\mathcal C}=\sum_{i=1}^F \mathcal C'_i\circ\bar{\mathcal C}\overset{(a)}{=}\bigg(\sum_{i=1}^F \mathcal C'_i\bigg)\circ\bar{\mathcal C},
\end{align} 
where $\C'_i+ \C'_j$ denotes the sum of subspaces $\C'_i$ and $\C'_j$, resulting in the set consisting of all elements $\bm c + \bm c'$ for any $\bm c\in\C'_i$ and $\bm c'\in\C'_j$, and where $(a)$ follows due to the fact that the Hadamard product is distributive over addition.

The symbols requested by the user are then obtained solving the system of linear equations defined by
\begin{align*}
  \bm H^{\tilde{\mathcal C}}\bm \rho_{j}=\bm H^{\tilde{\mathcal C}}\left(\begin{matrix}
      o^{(1)}_j
      \\
      o^{(2)}_j
      \\
      \vdots
      \\
      o^{(n)}_j
    \end{matrix}\right).
\end{align*}




\subsection{Privacy}

For the retrieval, we require $\tilde{\mathcal C}$ to be  a valid code, i.e., it must have a code rate strictly less than $1$. For a given number of colluding SBSs $T$, the combination of conditions on $\bar{\mathcal C}$ and $\tilde{\mathcal C}$ restricts the choice for the underlying storage codes $\{\mathcal C_i\}$. In the following theorem, we present a family of MDS codes, namely generalized Reed-Solomon (GRS) codes, that work with the protocol. A GRS code $\C$ over $\GF(q)$ of length $n$ and dimension $k$  is a weighted polynomial evaluation code of degree $k$ defined by some weighting vector $\bm v =(v_1,\ldots,v_n) \in (\GF(q)^{\times})^n$ and an evaluation vector $\boldsymbol{\kappa} = (\kappa_1,\ldots,\kappa_n) \in (\GF(q)^{\times})^n$ satisfying $\kappa_i \neq \kappa_j$ for all $i \neq j$ \cite[Ch.~5]{HuffmanPless10_1}.  In the sequel, we refer to $(n,k,\bm v,\boldsymbol{\kappa})$ as the parameters of a GRS code $\C$.

\begin{lemma}
\label{lem:GRS}
Given an $(n,\kmax, \bm v,\boldsymbol{\kappa})$ GRS code $\Cmax$, for all $k<\kmax$, there exists an $(n,k,\bm v,\boldsymbol{\kappa})$ GRS code that is a subcode of $\Cmax$.
\end{lemma}
\begin{IEEEproof}
The canonical generator matrix  for an $(n,\kmax,\bm v,\boldsymbol{\kappa})$ GRS code $\Cmax$  is given by
\begin{align}
\label{eq:GRS_gen}
\begin{pmatrix}
1 & 1 & \dots & 1 \\
\kappa_1 & \kappa_2 &  \dots & \kappa_n \\
\vdots & \vdots & \dots & \vdots  \\
\kappa_1^{\kmax-1} & \kappa_2^{\kmax-1} &  \dots & \kappa_n^{\kmax-1} \end{pmatrix} \begin{pmatrix}
v_1 & 0 & \dots & 0 \\
0 & v_2 & \dots & 0 \\
\vdots & \vdots & \dots & \vdots  \\
0 & 0 & \dots & v_n \end{pmatrix}.
\end{align}
Clearly, taking the first $k$ rows of the leftmost matrix of \eqref{eq:GRS_gen} and multiplying it with the rightmost diagonal matrix generates an $(n,k)$ subcode of $\Cmax$ which by itself is an $(n,k,\bm v,\boldsymbol{\kappa})$ GRS code. 
Thus, GRS codes are naturally nested, and the result follows.
\end{IEEEproof}

\begin{theorem}
\label{th:CachingScheme}
Let $\CMDS$ be a caching scheme with GRS codes $\{\C_i\}$ of parameters $(\NSBS,k_i,\bm v,(\kappa_1,\ldots,\kappa_{\NSBS}))$ and let $\C'_i$ be the $(n,k_i)$ code obtained by puncturing $\C_i$. Also,  let  $\bar{\mathcal C}$  be an $(n,T,\bar{\bm v} ,(\kappa_1,\ldots,\kappa_n))$ GRS code. Then, for $\beta=\Gamma=n-(k_{\mathsf{max}}+T-1)$ and $d=k_{\mathsf{max}}$, the protocol achieves PIR against $T$ colluding SBSs.
\end{theorem}

\ifnocomment
\begin{IEEEproof}
The proof is given in the appendix.
\end{IEEEproof}
\fi

Note that the retrieval code $\bar{\C}$ depends on the $n$ SBSs within visibility that are contacted by the user through its evaluation vector. Finally, we remark that, with some slight modifications, the proposed protocol can be adapted to work with non-MDS codes. 

\subsection{Example}

\begin{figure}
	\centering
	\includegraphics[width=\columnwidth]{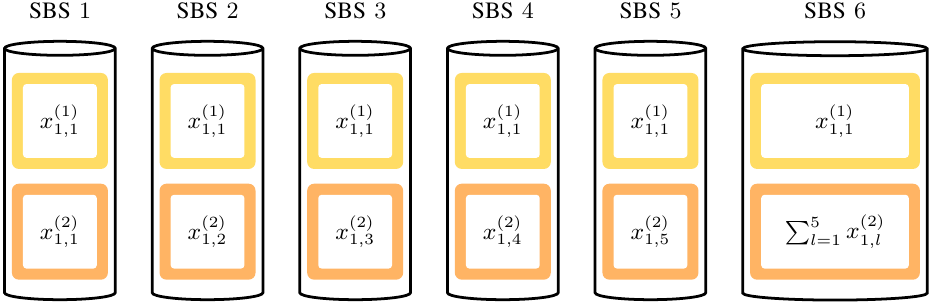}
	\caption{Wireless caching scenario in which there are $\NSBS=6$ SBSs. The SBSs store $F=2$ files,  $\bm X^{(1)}=(x^{(1)}_{1,1})\in\GF(2^5)^{1\times1}$ and $\bm X^{(2)}=(x^{(2)}_{1,1},x^{(2)}_{1,2},x^{(2)}_{1,3},x^{(2)}_{1,4},x^{(2)}_{1,5})\in\GF(2)^{1\times5}$, of $\beta L=5$ bits each. The first file $\bm X^{(1)}$ is encoded using an $(\NSBS=6,k_1=1)$ binary repetition code $\C_1$, while the second file $\bm X^{(2)}$ is encoded using an $(\NSBS=6,k_2=5)$ binary single parity-check code $\C_2$. }
	\label{Fig: Example}
\end{figure}

As an example, consider the case of $F=2$ files, $\bm X^{(1)}$ and $\bm X^{(2)}$, both of size $\beta L$ bits. The first file $\bm X^{(1)}$ is stored in the SBSs according to \cref{Fig: Example}  using an $(\NSBS=6,k_1=1)$ binary repetition code $\C_1$. Similarly, the second file $\bm X^{(2)}$ is stored (again according to \cref{Fig: Example}) using an $(\NSBS=6,k_2=5)$ binary single parity-check code $\C_2$. 
Assume $n=\NSBS=6$ (i.e., no puncturing) and that  none of the SBSs collude, i.e., $T=1$. Furthermore, we assume that the user wants to retrieve $\bm X^{(1)}$ and is able to contact $b=n=6$ \acp{SBS} (i.e., we consider the extreme case where the user is not contacting the MBS). According to \cref{th:CachingScheme}, we can choose $\beta=\Gamma=n-(k_{\mathsf{max}}+T-1)= 6-(5+1-1)=1$ and $d=k_{\mathsf{max}}=5$. Finally,  we choose $\bar{\mathcal C}$ as an $(n=6,T=1)$ binary repetition code. 

According to \eqref{Eq: Ctilde}, the retrieval code $\tilde{\mathcal C}= (\mathcal C_1 + \mathcal C_2) \circ  \bar{\mathcal C} = \C_1 + \C_2 = \C_2$ and can be generated by
\begin{align*}
	\bm G^{\tilde{\mathcal C}}= 
	\bm G^{\mathcal C_2}=\left(\begin{matrix}
		1&0&0&0&0&1\\
		0&1&0&0&0&1\\
		0&0&1&0&0&1\\
		0&0&0&1&0&1\\
		0&0&0&0&1&1
	\end{matrix}\right).
\end{align*}
Moreover, let
\begin{align*}
	\hat{\bm E}=\left(\begin{matrix}
		1&0&0&0&0&0\\
		0&1&0&0&0&0\\
		0&0&1&0&0&0\\
		0&0&0&1&0&0\\
		0&0&0&0&1&0
	\end{matrix}\right)\,\,\,\, \text{and}\,\,\,\, \mathcal I_1=\{1,2,3,4,5\},
\end{align*}
where $\mathcal I_1$ is an information set of $\Cmax=\C_2$ (the submatrix $\bm G^{\mathcal C_2}|_{\mathcal{I}_1}$ has rank $k_2=5$).  
Note that $\hat{\bm E}$ satisfies all three conditions $\mathsf{C1}$--$\mathsf{C3}$ and has column weight profile $(1,1,1,1,1,0)=(|\mathcal{F}_1|,\ldots,|\mathcal{F}_6|)$.

\emph{Query Construction.} The user generates $\beta F=2$ codewords $\bar{\bm c}^{(1)}_1$ and $\bar{\bm c}^{(2)}_1$ independently and uniformly at random from $\bar{\mathcal C}$. Without loss of generality, let $\bar{\bm c}^{(1)}_1=\bar{\bm c}^{(2)}_1=(1,\ldots,1)$. 
Next, the $n=6$  subqueries $q_1^{(l)}$, $l=1,\ldots,6$, are constructed according to  \cref{Eq: CollQuery_designa}, \cref{Eq: CollQuery_design} as
\begin{align*}
\bm q_1^{(l)} = \begin{cases}
\mathring{\bm c}_l+(1,0) & \text{if $l=1$}, \\
\mathring{\bm c}_l+(0,0) & \text{otherwise}, \end{cases}
%
%
\end{align*}
where $\mathring{\bm c}_l$ is defined in \eqref{eq:vector_circle}.

\emph{File Retrieval.} Consider the $n=6$ subresponses $r_1^{(l)}$, $l=1,\ldots,6$.  Then, according to \eqref{Eq: PIR_coll_response}, 
\begin{align*}
  \bm \rho_{1} =\left(\begin{matrix}
      r_{1}^{(1)}\\
      r_{1}^{(2)}\\
      r_{1}^{(3)}\\
      r_{1}^{(4)}\\
      r_{1}^{(5)}\\
      r_{1}^{(6)}
    \end{matrix}\right) &=
  \underbrace{\left(\begin{matrix}
      \bar c_{1,1}^{(1)}c_{1,1}^{(1)}\\
      \bar c_{1,2}^{(1)}c_{1,2}^{(1)}\\
      \vdots\\
      \bar c_{1,6}^{(1)}c_{1,6}^{(1)}\\
    \end{matrix}\right)}_{\substack{\in\, \left\{  \bm x \in  (\GF(2^5))^n \colon \right. \\ \left. \bm H^{\C'_1\circ\,\bar{\mathcal C}} \bm x = \bm 0  \right\}  }}
    +\underbrace{
  \left(\begin{matrix}
      \bar c_{1,1}^{(2)}c_{1,1}^{(2)}\\
      \bar c_{1,2}^{(2)}c_{1,2}^{(2)}\\
      \vdots\\
      \bar c_{1,6}^{(2)}c_{1,6}^{(2)}\\
    \end{matrix}\right)}_{\substack{\in\, \left\{  \bm x \in  (\GF(2^5))^n \colon \right. \\ \left. \bm H^{\C'_2\circ\,\bar{\mathcal C}} \bm x = \bm 0  \right\}  }}
  +\left(\begin{matrix}
      o_1^{(1)}\\
      o_1^{(2)}\\
      \vdots\\
      o_{1}^{(6)}
  \end{matrix}\right) \\
  &=\left(\begin{matrix}
      x_{1,1}^{(1)}\\
      x_{1,1}^{(1)}\\
      x_{1,1}^{(1)}\\
      x_{1,1}^{(1)}\\
      x_{1,1}^{(1)}\\
      x_{1,1}^{(1)}\\
    \end{matrix}\right)+\left(\begin{matrix}
      x_{1,1}^{(2)}\\
      x_{1,2}^{(2)}\\
      x_{1,3}^{(2)}\\
      x_{1,4}^{(2)}\\
      x_{1,5}^{(2)}\\
      \sum_{l=1}^5x_{1,l}^{(2)}\\
    \end{matrix}\right)+\left(\begin{matrix}
      x_{1,1}^{(1)}\\
      0\\
      0\\
      0\\
      0\\
      0\\
    \end{matrix}\right),
\end{align*}
and the code symbol $x_{1,1}^{(1)}$ of the file $\bm X^{(1)}$ is recovered from
\begin{align*}
	 \bm H^{\tilde{\mathcal C}}\bm \rho_{1}=\left(\begin{matrix}1&1&1&1&1&1\end{matrix}\right)\left(\begin{matrix}
      x_{1,1}^{(1)}\\
      0\\
      0\\
      0\\
      0\\
      0\\
    \end{matrix}\right)=x^{(1)}_{1,1}.
\end{align*}
Note that in order to retain privacy across the two files of the library, we need to send $d=k_{\mathsf{max}}=5$ subqueries to each SBS, thus generating $5$ subresponses from each SBS (even if the first file can be recovered from the $n=6$ subresponses $r_1^{(l)}$, $l=1,\ldots,6$).

\section{Backhaul Rate Analysis: No PIR Case}
\label{sec:CAnPIR}

In this section, we derive the backhaul rate for the proposed caching scheme for the case of no PIR, i.e., the conventional caching scenario where PIR is not required.
\begin{proposition}
The average backhaul rate for the caching scheme $\CMDS$ in Section~\ref{sec:SystemModel} for the case of no PIR is
\begin{align}
\label{eq:noPIR}
&\RnPIR\nonumber\\
&= \sum_{i=1}^F  p_i \lceil \mu_i \rceil \sum_{b=0}^{\NSBS}\gamma_b  \max\left(0,1/\mu_i-b\right)\mu_i + \sum_{i=1}^Fp_i \lfloor 1-\mu_i\rfloor .
\end{align}
\end{proposition}
\ifnocomment
\begin{IEEEproof}
To download file $\bXi{i}$, if the user is in communication range of a number of \acp{SBS}, $b$, larger than or equal to $1/\mu_i$, the user can retrieve the file from the \acp{SBS} and there is no contribution to the backhaul rate. Otherwise, if $b<1/\mu_i$, the user retrieves a fraction $L/k_i=L\mu_i$ of the file from each of the $b$ \acp{SBS}, i.e., a total of $b\beta L\mu_i$ bits, and downloads the remaining $(1/\mu_i-b) \beta L\mu_i$ bits from the MBS. Averaging over $\bm \gamma$ and $\bm p$ (for the files cached) and normalizing by the file size $\beta L$, the contribution to the backhaul rate of the retrieval of files that are cached in the \acp{SBS} is 
\begin{align}
\label{eq:BhAnPIR}
\sum_{i=1}^F  p_i \lceil \mu_i \rceil \sum_{b=0}^{\NSBS}\gamma_b  \max\left(0,1/\mu_i-b\right)\mu_i.
\end{align}
On the other hand, the files that are not cached are retrieved completely from the MBS, and their contribution to the backhaul rate is 
\begin{align}
\label{eq:BhBnPIR}
\sum_{i=1}^Fp_i \lfloor 1-\mu_i\rfloor.
\end{align}
Combining \eqref{eq:BhAnPIR} and \eqref{eq:BhBnPIR} completes the proof.
\end{IEEEproof}
\fi

We denote by $\RnPIR^*$ the maximum PIR rate resulting from the optimization of the content placement.
$\RnPIR^*$ can be obtained solving the following optimization problem,
\begin{align*}
\RnPIR^*&= \min_{\mu_i\in\mathcal{M}'}~~\sum_{i=1}^F  p_i\lceil \mu_i \rceil\sum_{b=0}^{\NSBS} \gamma_b   \max\big(0,1/\mu_i-b\big) \mu_i\nonumber\\ 
&~~~~~~~~~~~~~+\sum_{i=1}^Fp_i \lfloor 1-\mu_i\rfloor\\
& \text{s.t.} \sum_{i=1}^F \mu_i \le M,\nonumber
\end{align*}
where $\mathcal M'=\mathcal M\cup \{1/\NSBS\}$, as $\mu_i=1/\NSBS$ is a valid value for the case where PIR is not required.

In the following lemma, we show that the proposed content placement is equivalent to the one in \cite{Bio15}, in the sense that it yields the same average backhaul rate.
\begin{lemma}
\label{lem:equiv}
The average backhaul rate given by \eqref{eq:noPIR} for the caching scheme $\CMDS$ in Section~\ref{sec:SystemModel}  is equal to the one given by the caching scheme in \cite{Bio15}, i.e., the two content placements are equivalent.
\end{lemma}
\ifnocomment
\begin{IEEEproof}
We can rewrite \eqref{eq:noPIR} using simple math as
\begin{align*}
&\RnPIR
\\
&=\sum_{i=1}^F  p_i\lceil \mu_i \rceil\sum_{b=0}^{\NSBS} \gamma_b   \max\big(0,1/\mu_i-b\big) \mu_i +\sum_{i=1}^Fp_i \lfloor 1-\mu_i\rfloor\\
&=\sum_{i=1}^F  p_i\lceil \mu_i \rceil\sum_{b=0}^{\NSBS} \gamma_b   \max\big(0,1-b\mu_i\big)  +\sum_{i=1}^Fp_i \lfloor 1-\mu_i\rfloor\\
&=\sum_{i=1}^F  p_i\lceil \mu_i \rceil\sum_{b=0}^{\NSBS} \gamma_b   \big(1-\min\big(1,b\mu_i\big)\big)  +\sum_{i=1}^Fp_i \lfloor 1-\mu_i\rfloor\\
&\stackrel{(a)}{=}\sum_{i=1}^F  p_i(\lceil \mu_i \rceil + \lfloor 1-\mu_i\rfloor)\sum_{b=0}^{\NSBS} \gamma_b   \big(1-\min\big(1,b\mu_i\big)\big)\\
&=\sum_{i=1}^F  p_i \sum_{b=0}^{\NSBS} \gamma_b   \big(1-\min\big(1,b\mu_i\big)\big), 
\end{align*}
which is the expression in \cite[eq.~(1)]{Bio15}. $(a)$ follows from the fact that we can write $p_i \lfloor 1-\mu_i\rfloor$ as  $p_i\lfloor 1-\mu_i\rfloor\sum_{b=0}^{\NSBS} \gamma_b   \big(1-\min\big(1,b\mu_i\big)\big)$. For $0<\mu_i\le 1$ both expressions are zero, while for $\mu_i=0$ both expressions boil down to $p_i$ as $p_i\lfloor 1-\mu_i\rfloor\sum_{b=0}^{\NSBS} \gamma_b   \big(1-\min\big(1,b\mu_i\big)\big)=p_i\sum_{b=0}^{\NSBS} \gamma_b$ and $\sum_{b=0}^{\NSBS} \gamma_b=1$.
\end{IEEEproof}
\fi

For popular content placement, i.e., the case where the $M$ most popular files are cached in all SBSs (this corresponds to caching the $M$ most popular files using an $(\NSBS,1)$ repetition code,  i.e., $\mu_i=1$ for $i\le M$ and $\mu_i=0$ for $i > M$), the backhaul rate is given by
\begin{align}
\RnPIRpop&=\gamma_0\sum_{i=1}^M p_i  + \sum_{i=M+1}^Fp_i.
\label{eq:nPIRpopular}
\end{align}

\section{Backhaul Rate Analysis: PIR Case}
\label{sec:CAPIR}

In this section, we derive the backhaul rate for the case of PIR (i.e., when the user wishes to download content privately) and we prove that uniform content placement (under the PIR protocol in Section~\ref{sec:PIRprotocol} with GRS codes) is optimal. The average backhaul rate is given in the following proposition.

\begin{proposition}
The average backhaul rate  for the caching scheme $\CMDS$ in Section~\ref{sec:SystemModel} (with GRS codes) for the PIR case is 
\begin{align}
\RPIR\nonumber=\;&\frac{\mumax}{\mu_{\min}(n-T+1)-1 }\sum_{i=1}^F  p_i \lceil \mu_i \rceil \sum_{b=0}^{n} \gamma_b  (n-b) \\
&+\sum_{i=1}^Fp_i \lfloor 1-\mu_i\rfloor .
\label{eq:cost_MBS}
\end{align}
\end{proposition}
\ifnocomment
\begin{IEEEproof}
To download file $\bXi{i}$, the user generates $n$ query matrices. If the user is in communication range of $b$ \acp{SBS}, it receives $b$ responses (one from each SBS). The responses to the remaining $n-b$ query matrices need to be downloaded from the MBS. Since each response consists of $d$ subresponses of size $L\mumax$ bits, the user downloads $(n-b)d L\mumax$ bits from the MBS. Averaging over $\bm \gamma$ and $\bm p$ (for the files cached) and normalizing by the file size $\beta L$, the contribution to the backhaul rate of the retrieval of files that are cached in the \acp{SBS} is 
\begin{align}
\label{eq:BhA}
\frac{1}{\beta}\sum_{i=1}^F  p_i \lceil \mu_i \rceil\sum_{b=0}^{n} \gamma_b   (n-b)d \mumax.
\end{align}
Now, using the fact that $\beta=\Gamma=n-(\kmax+T-1)=\frac{\mu_{\mathsf{min}}(n-T+1)-1}{\mu_{\mathsf{min}}}$ and $d=k_{\mathsf{max}}=1/\mumin$ (see Theorem~\ref{th:CachingScheme}), we can rewrite \eqref{eq:BhA} as
\begin{align}
\label{eq:BhB}
\frac{\mumax}{\mu_{\min}(n-T+1)-1 }\sum_{i=1}^F  p_i \lceil \mu_i \rceil \sum_{b=0}^{n} \gamma_b  (n-b).
\end{align}

On the other hand, the files that are not cached are retrieved completely from the MBS, and their contribution to the backhaul rate is (as for the no PIR case)
\begin{align}
\label{eq:BhC}
\sum_{i=1}^Fp_i \lfloor 1-\mu_i\rfloor.
\end{align}
Combining \eqref{eq:BhB} and \eqref{eq:BhC} completes the proof.
\end{IEEEproof}
\fi

\subsection{Optimal Content Placement}

Let $\RPIR^*$ be the maximum PIR rate resulting from the optimization of the content placement.
$\RPIR^*$ can be obtained solving the following optimization problem,
\begin{align}
\label{eq:CBSModifiedopt}
\RPIR^*
&=\underset{\substack{\mu_i\in\mathcal{M}\\ n\in\mathcal A}}\min~
\frac{\mumax}{\mu_{\min}(n-T+1)-1 }\sum_{i=1}^F  p_i \lceil \mu_i \rceil \sum_{b=0}^{n} \gamma_b (n-b) \nonumber\\
&~~~~~~~~~~~+\sum_{i=1}^Fp_i \lfloor 1-\mu_i\rfloor\\
& \text{s.t.} \sum_{i=1}^F \mu_i \le M\;\text{and}\;\kmin\mid k_i, \nonumber
\end{align}
where $\mathcal A=\{1/\mumin+T,\ldots,\NSBS\}$ and the minimum value that $n$ can take on, i.e., $1/\mumin+T$, comes from the fact that $\mu_{\min}(n-T+1)-1 $ has to be positive.

\begin{lemma} \label{lem:min_eq_max}
Uniform content allocation, i.e., $\mu_i=\mu$ for all files that are cached, is optimal. Furthermore, the optimal number of files to cache is the maximum possible, i.e., $\mu_i=\mu$  for $i\le\min(M/\mu,F)$.
\end{lemma}
\ifnocomment
\begin{IEEEproof}
We first prove the first part of the lemma. We need to show that either the optimal solution to the optimization problem in \eqref{eq:CBSModifiedopt} is the all-zero vector $\boldsymbol{\mu} = (\mu_1,\ldots,\mu_F)=(0,\ldots,0)$, or there exists a nonzero optimal solution $\boldsymbol{\mu} = (\mu_1,\ldots,\mu_F)$ for which $\mumax = \mumin$.
Consider the second case, and let $\boldsymbol{\mu}$ denote any nonzero feasible solution to  \eqref{eq:CBSModifiedopt}, i.e., a nonzero solution that satisfies the cache size constraint. Furthermore, let $\boldsymbol{\mu}'=(\mu'_1,\ldots,\mu'_F)$ denote the length-$F$ vector obtained from $\boldsymbol{\mu}$ as $\mu'_i=\mumin$ for $\mu_i\neq 0$ and $\mu'_i=0$ otherwise. Clearly, $\boldsymbol{\mu}'$ satisfies the cache size constraint as well. Note that $\mumax'=\mumin'=\mumin$. Thus,
\begin{align*}
 \frac{\mumax'}{\mumin'(n-T+1)-1} &= \frac{\mumin}{\mumin(n-T+1)-1} \\ &\leq \frac{\mumax}{\mumin(n-T+1)-1}.
\end{align*} 
  Furthermore, since both the double summation in the first term of the objective function in  \eqref{eq:CBSModifiedopt} and the second term in \eqref{eq:CBSModifiedopt} only depend on the support of $\boldsymbol{\mu}$, it follows that the value of the objective function for $\boldsymbol{\mu}'$ is smaller than or equal to the value of the objective function for $\boldsymbol{\mu}$. Thus, for any nonzero feasible solution $\bm \mu$ there exists another at least as good  nonzero feasible solution $\bm\mu'$ for which all nonzero entries are the same (i.e., $\mumin'=\mumax'=\mu$), and the result follows by applying the above procedure to a (nonzero) optimal solution to  \eqref{eq:CBSModifiedopt}.
 
We now prove the second part of the lemma. Caching a file helps in reducing the backhaul rate if
\begin{align}
\label{eq:cond}
\frac{\mu}{\mu(n-T+1)-1 } \sum_{b=0}^{n} \gamma_b  (n-b)< 1,
\end{align}
for some $n\in\mathcal A$ and $\mu \in \mathcal{M}$. 
This is independent of the file index $i$. Thus, if the optimal solution is to cache at least one file ($\bm\mu\neq \bm 0$), \eqref{eq:cond} is met for some $n\in\mathcal A$ and caching other files
(as many files as permitted up to the cache size constraint, with decreasing order of popularity) is optimal as it further reduces the backhaul rate.
\end{IEEEproof}
\fi

Following Lemma~\ref{lem:min_eq_max}, the optimization problem in \eqref{eq:CBSModifiedopt} can be rewritten as
\begin{align}
\RPIR^*  = \underset{\substack{\mu\in\mathcal{M}\\ n\in\mathcal A}}\min\; &\frac{\mu}{\mu(n-T+1)-1 } \sum_{i=1}^{\min(M/\mu,F)}  p_i \sum_{b=0}^{n} \gamma_b  (n-b)\nonumber\\
& +\sum_{i=M/\mu+1}^Fp_i .
\label{eq:optimFixedmuMod}
\end{align}

\subsection{Popular Content Placement}
For popular content placement,  the backhaul rate is given by
\begin{align}
\RPIR^\mathsf{pop}=\min_{n\in\mathcal A}\;&\frac{1}{n-T } \sum_{i=1}^{M}  p_i \sum_{b=0}^{n} \gamma_b  (n-b) 
+ \sum_{i=M+1}^Fp_i.
\label{eq:popularMod}
\end{align}
Note that the optimization over $n$ is still required.

\section{Weighted Communication Rate}

So far, we have considered only the backhaul rate. However, it might also be desirable to limit the communication rate from SBSs to the user. We thus consider the weighted communication rate, $\CPIR$, defined as\footnote{For the case of no PIR, a linear scalarization of the MBS and SBS download delays was considered in \cite{Sha13}. The communication
rate is directly related to the download delay.}
\begin{align*}
\CPIR =  \RPIR + \theta\DPIR,
\end{align*}
where $\DPIR$ is the average communication rate (normalized by the file size $\beta L$) from the \acp{SBS}, and $\theta$ is a weighting parameter. We consider $\theta \le 1$, stemming from the fact that the bottleneck is the backhaul. Note that minimizing the average backhaul rate corresponds to $\theta=0$.

\begin{proposition}
The average communication rate from the SBSs for the caching scheme $\CMDS$ in Section~\ref{sec:SystemModel} (with GRS codes) for the PIR case is 
\begin{align}
\DPIR=
\frac{\mumax}{\mu_{\min}(n-T+1)-1 }\sum_{b=0}^{n} \tilde{\gamma}_b  b ,
\label{eq:SBSModified}
\end{align}
where $\tilde{\gamma}_b=\gamma_b$ for $b<n$ and $\tilde{\gamma}_{n}=\sum_{b=n}^{\NSBS}\gamma_b$.
\end{proposition}
\begin{IEEEproof}
To ensure privacy, the user needs to download data from the SBSs within visibility regardless whether the requested file is cached or not. This is in contrast to the case of no PIR. Note that, if the user queries the SBSs only in the case the requested file is cached, then the spy SBSs would infer that the user is interested in one of the files cached, thus gaining some information about the file requested. In other words, the user sends dummy queries and downloads data that is useless for the retrieval of the file but is necessary to achieve privacy. The user receives $b$ responses from the $b$ SBSs within communication range, each of size $dL\mumax$ bits. Let $\tilde{\gamma}_{b}$ denote the probability to receive responses from $b$ SBSs. For $b<n$, $\tilde{\gamma}_{b}$ is equal to the probability that $b$ SBSs are within communication range, i.e., $\tilde \gamma_b=\gamma_b$. On the other hand, the probability to receive responses from $n$ SBSs, $\tilde{\gamma}_{n}$, is the probability that at least $n$ SBSs are within communication range, i.e., $\tilde{\gamma}_{n}=\sum_{b=n}^{\NSBS}\gamma_b$. Averaging over $\tilde{\bm \gamma}$ and $\bm p$ (for all files, cached and not cached) and normalizing by the file size $\beta L$, the contribution to the communication rate  of the retrieval of a file from the SBSs is
\begin{align}
\label{eq:CA}
\frac{1}{\beta}\sum_{i=1}^F  p_i \sum_{b=0}^{n} \tilde\gamma_b   b d \mumax.
\end{align}
Now, using the fact that $\beta=\Gamma=n-(\kmax+T-1)=\frac{\mu_{\mathsf{min}}(n-T+1)-1}{\mu_{\mathsf{min}}}$ and $d=k_{\mathsf{max}}=1/\mumin$ (see Theorem~\ref{th:CachingScheme}), we can rewrite \eqref{eq:CA} as \eqref{eq:SBSModified}.
\end{IEEEproof}

The corresponding optimization problem is
\begin{align}
\label{eq:Cmodopt}
\CPIR^*&= \underset{\substack{\mu_i\in\mathcal{M}\\ n\in\mathcal A}}\min \;\RPIR + \theta\DPIR\\
& \text{s.t.} \sum_{i=1}^F \mu_i \le M\;\text{and}\;\kmin\mid k_i, \nonumber
\end{align}
where $\RPIR$ is given in \eqref{eq:cost_MBS}.

\begin{lemma} \label{lem:min_eq_max_CSBS}
Uniform content allocation, i.e., $\mu_i=\mu$ for all files that are cached, is optimal. Furthermore, the optimal number of files to cache is the maximum possible, i.e., $\mu_i=\mu$  for $i\le\min(M/\mu,F)$.
\end{lemma}
\begin{IEEEproof}
The proof of Lemma~\ref{lem:min_eq_max} applies to both terms in \eqref{eq:Cmodopt} and the result follows.
\end{IEEEproof}

Following Lemma~\ref{lem:min_eq_max_CSBS}, the optimization problem in \eqref{eq:Cmodopt} can be rewritten as
\begin{align}
\CPIR^*  &= \underset{\substack{\mu\in\mathcal{M}\\ n\in\mathcal A}}\min~~\frac{\mu}{\mu(n-T+1)-1 } \sum_{i=1}^{\min(M/\mu,F)}  p_i \sum_{b=0}^{n} \gamma_b  (n-b)\nonumber\\
& ~~~~~~~~~~~+\sum_{i=M/\mu+1}^Fp_i  + \theta\frac{\mu}{\mu(n-T+1)-1 }\sum_{b=0}^{n} \tilde{\gamma}_b  b .
\label{eq:optC}
\end{align}

\section{Numerical Results}
 
\begin{figure}[!t]
\begin{centering}
\includegraphics[width=\columnwidth]{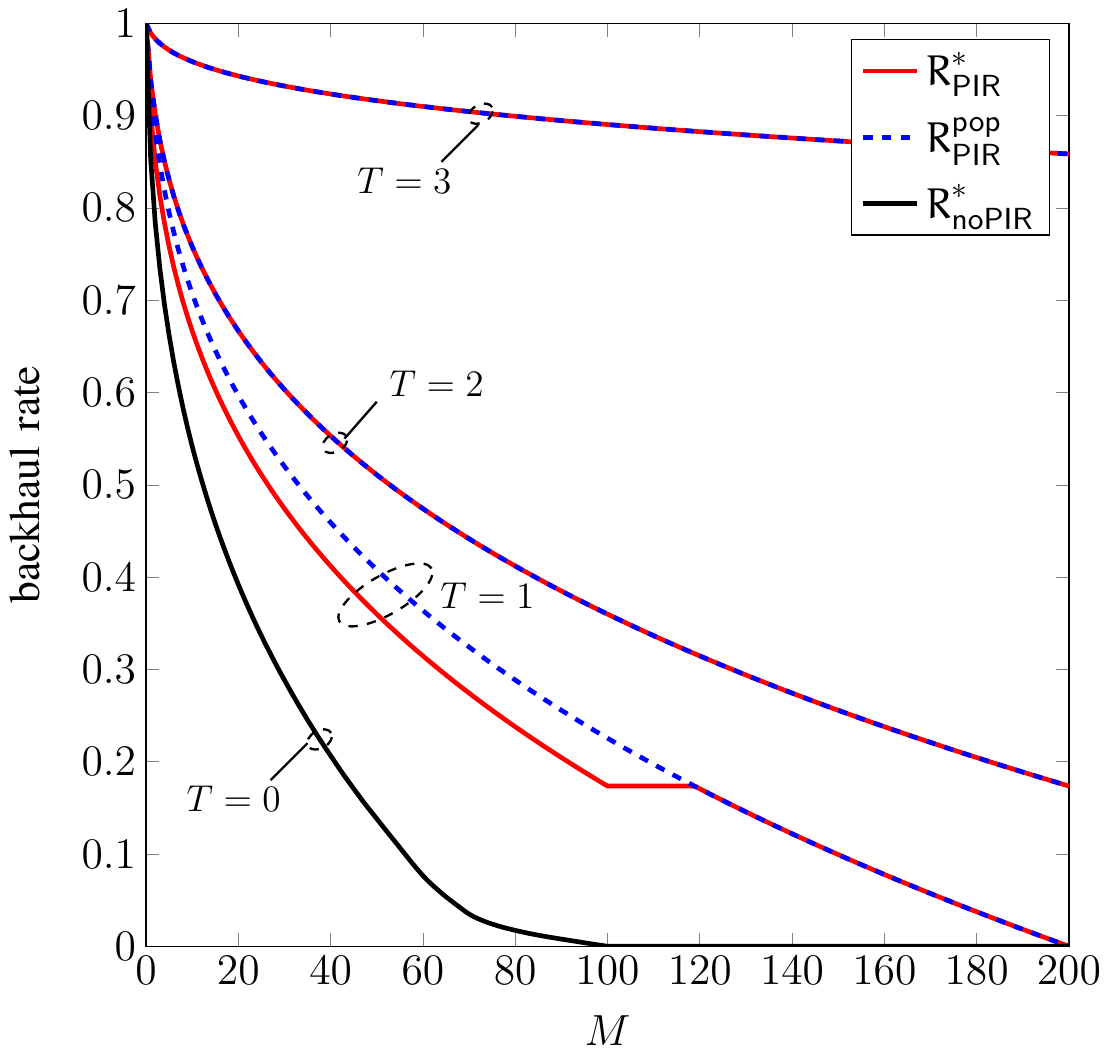}
\caption{Backhaul rate as a function of the cache size constraint $M$ for a system with $F = 200$ files, $\NSBS=316$, and $\alpha=0.7$. }
\label{fig:optim_diff_M}
\par\end{centering}
\end{figure}

\begin{figure}[!t]
\begin{centering}
\includegraphics[width=\columnwidth]{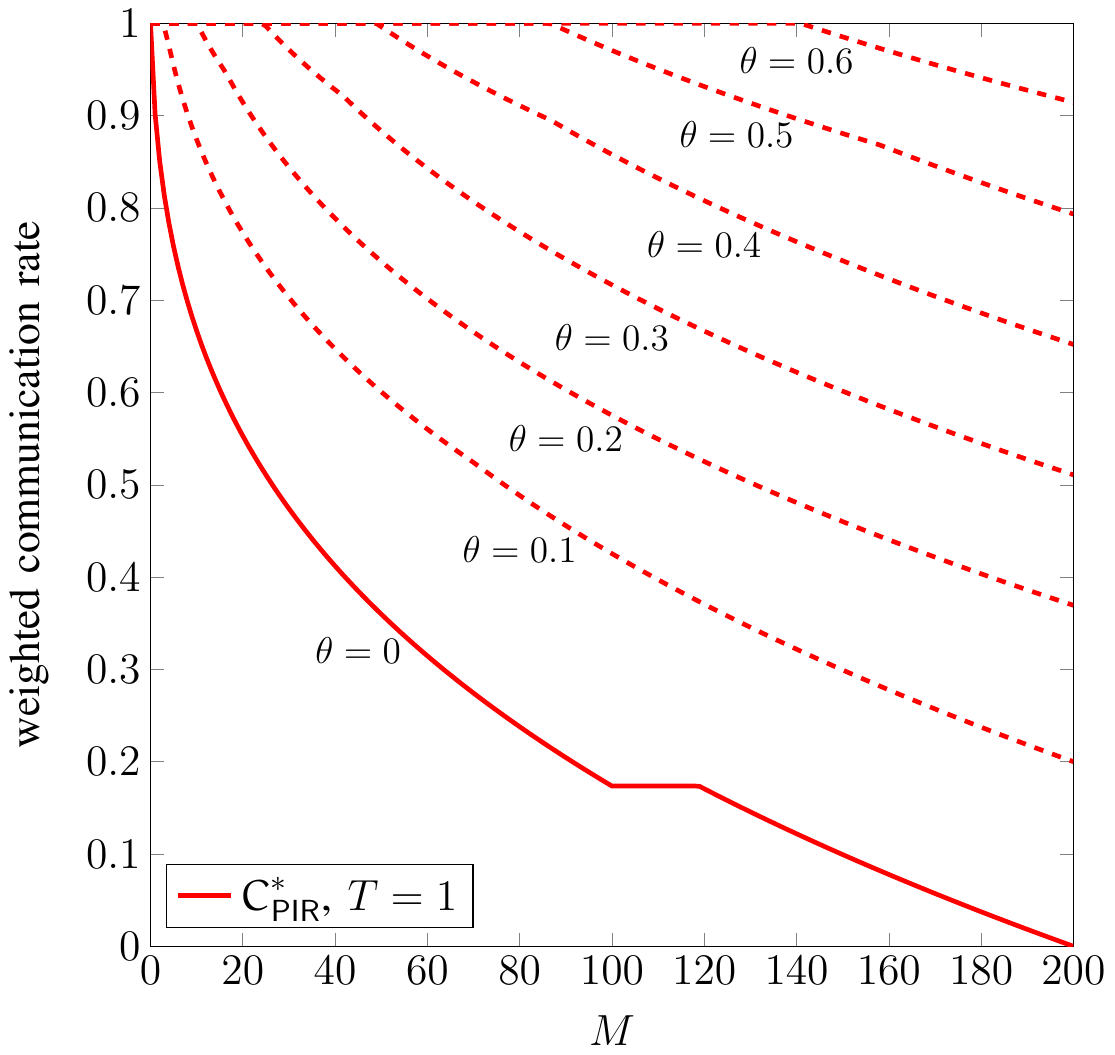}
\caption{Optimized weighted communication rate as a function of the cache size constraint $M$ for a system with $T=1$ spy SBS, $F = 200$ files, $\NSBS=316$, $\alpha=0.7$, and several values of $\theta$.}
\label{fig:RWeighted}
\par\end{centering}
\end{figure}

For the numerical results in this section, we assume that the files popularity distribution $\bp$ follows the Zipf law \cite{Bre99}, i.e., the popularity of file $\bXi{i}$ is
\begin{align*}
	p_i=\frac{1/i^\alpha}{\sum_\ell 1/\ell^\alpha},
\end{align*}
where $\alpha\in[0.5,1.5]$ is the skewness factor \cite{Bio15} and by definition $p_1\ge p_2\ge \ldots \ge p_F$. In Figs.~\ref{fig:optim_diff_M} and \ref{fig:RWeighted}, we consider a network topology where SBSs are deployed over a macro-cell of radius $D$ meters according to a regular grid with distance $d$ meters between them \cite{Bio15,Sha13}. Each SBS has a communication radius of $r$ meters. Let $\mathcal{R}_b$ be the area where a user can be served by $b$ \acp{SBS}. Then, assuming that the users are uniformly distributed over the macro-cell area with density $\phi$ users per square meter, the probability that 
  a user is in communication range of $b$ \acp{SBS} can be calculated as in \cite{Bio15}
\begin{equation*}
\gamma_b = \frac{\phi \mathcal{R}_{b}}{\phi\sum_{a=1}^{N_{\mathsf{max}}}  \mathcal{R}_a},
\end{equation*}
where the areas $\mathcal R_b$ can be easily obtained by simple geometrical evaluations, and $N_{\mathsf{max}}$ is the maximum number of SBSs within communication range of a user.

For the results in Figs.~\ref{fig:optim_diff_M} and \ref{fig:RWeighted}, the system parameters (taken from  \cite{Bio15}) are $D=500$ meters, which results in $\NSBS=316$ over the macro-cell area, $F=200$ files, $\alpha=0.7$, and $r=60$ meters. This results in $\bm\gamma=(0,0,0.1736,0.5113,0.3151,0,\ldots,0)$, i.e., the maximum number of SBSs in visibility of a user is $N_{\mathsf{max}}=4$.

In Fig.~\ref{fig:optim_diff_M}, we plot the optimized backhaul rate $\RPIR^*$ (red, solid lines) according to \eqref{eq:optimFixedmuMod} as a function of the cache size constraint $M$ for the noncolluding case ($T=1$) and $T=2$ and $T=3$ colluding SBSs. The curves in Fig.~\ref{fig:optim_diff_M} should be interpreted as the minimum backhaul rate that is necessary in order to achieve privacy against $T$ spy SBSs out of the $n$ SBSs that are contacted by the user. For the particular system parameters considered, the optimal value of $n$ is $3$ for $T=1$ and $T=2$, and all values of $M$, i.e., the scheme yields privacy against $T$ spy SBSs out of the $n=3$ SBSs contacted. For $T=3$ the optimal value of $n$ is $4$ for all values of $M$, and thus the scheme yields privacy against $3$ spy SBSs out of $n=4$ SBSs. We also plot the optimized backhaul rate $\RnPIR^*$ for the case of no PIR.\footnote{The curve $\RnPIR^*$ in the figure is identical to that in \cite[Fig.~4]{Bio15}. As proved in Lemma~\ref{lem:equiv}, while the proposed content placement is different from the one in \cite{Bio15}, they are equivalent in terms of average backhaul rate.}  As can be seen in the figure, caching helps in significantly reducing the backhaul rate for $T=1$ and $T=2$. For $T=3$ caching also helps in reducing the backhaul rate, but the reduction is smaller. Also, as expected, compared to the case of no PIR ($\RnPIR^*$, black, solid line) achieving privacy requires a higher backhaul rate. The required backhaul rate increases with the number of colluding SBSs $T$.

For $M\ge 100$ and no PIR, the backhaul rate is zero, as all files can be downloaded from the SBSs. Indeed, for $M=100$, we can select $k_i=2~\forall i$ and cache one coded symbol from each stripe of each file in each SBS (thus satisfying the constraint  $\sum_{i=1}^{F}\mu_i\le M$ as $\sum_{i=1}^{200} \mu_i = \sum_{i=1}^{200} 1/k_i =  \sum_{i=1}^{200} 0.5 =  100 $). Since for no PIR  to retrieve each stripe of a file it is enough to download $2$ symbols  from each stripe of the file (due to the MDS property) and according to $\bm\gamma$ at least $2$ SBSs are within range, for $M=100$ (and hence for $M>100$ as well) the user can always retrieve the file from the SBSs and the backhaul rate is zero. For the case of PIR and $T=1$, on the other hand, the required backhaul rate is positive unless all complete files can be cached in all SBSs, i.e., $M=F$. For $T=2$ and $T=3$, even for $M=F$ the backhaul rate is not zero. This is because in this case the user needs to receive $n=3$ and $n=4$ responses $\brl{l}$, $l=1,\ldots,n$, respectively (from the SBSs or the MBS).  However, 
for the considered system parameters the probability that the user has $b\ge 3$ SBSs within range is not one, thus the user always needs to download data from the MBS to recover the file and the backhaul rate is positive.


For comparison purposes, in the figure we also plot the backhaul rate for the case of popular content placement $\RPIRpop$ in \eqref{eq:popularMod} (blue, dashed lines). In this case, the optimal value of $n$ is $2$, $3$, and $4$ for $T=1$, $T=2$, and $T=3$, respectively. We remark that the curve $\RPIRpop$ for $T=1$ overlaps with the curve $\RnPIRpop$.
This is due to the fact that for $T=1$, $n=2$, and $\gamma_0=\gamma_1=0$, $\RPIRpop$ in \eqref{eq:popularMod} boils down to $\sum_{M+1}^Fp_i$, which is $\RnPIRpop$ in \eqref{eq:nPIRpopular}. However, for the general case, i.e., other $\bm \gamma$, $\RPIRpop$ and $\RnPIRpop$ may differ. As already shown in \cite{Bio15}, for no PIR the optimized content placement yields significantly lower backhaul rate than popular content placement. For the PIR case and $T=1$, up to $M=118$ the optimized content placement also yields some performance gains with respect to popular content placement, albeit not as significant as for the case of no PIR. Interestingly, as shown in the figure, for $M\ge 119$, PIR popular content placement is optimal. Furthermore, as shown in the figure, for $T=2$ and $T=3$ popular content placement is optimal for all $M$. 
\begin{figure}[!t]
\begin{centering}
\includegraphics[width=\columnwidth]{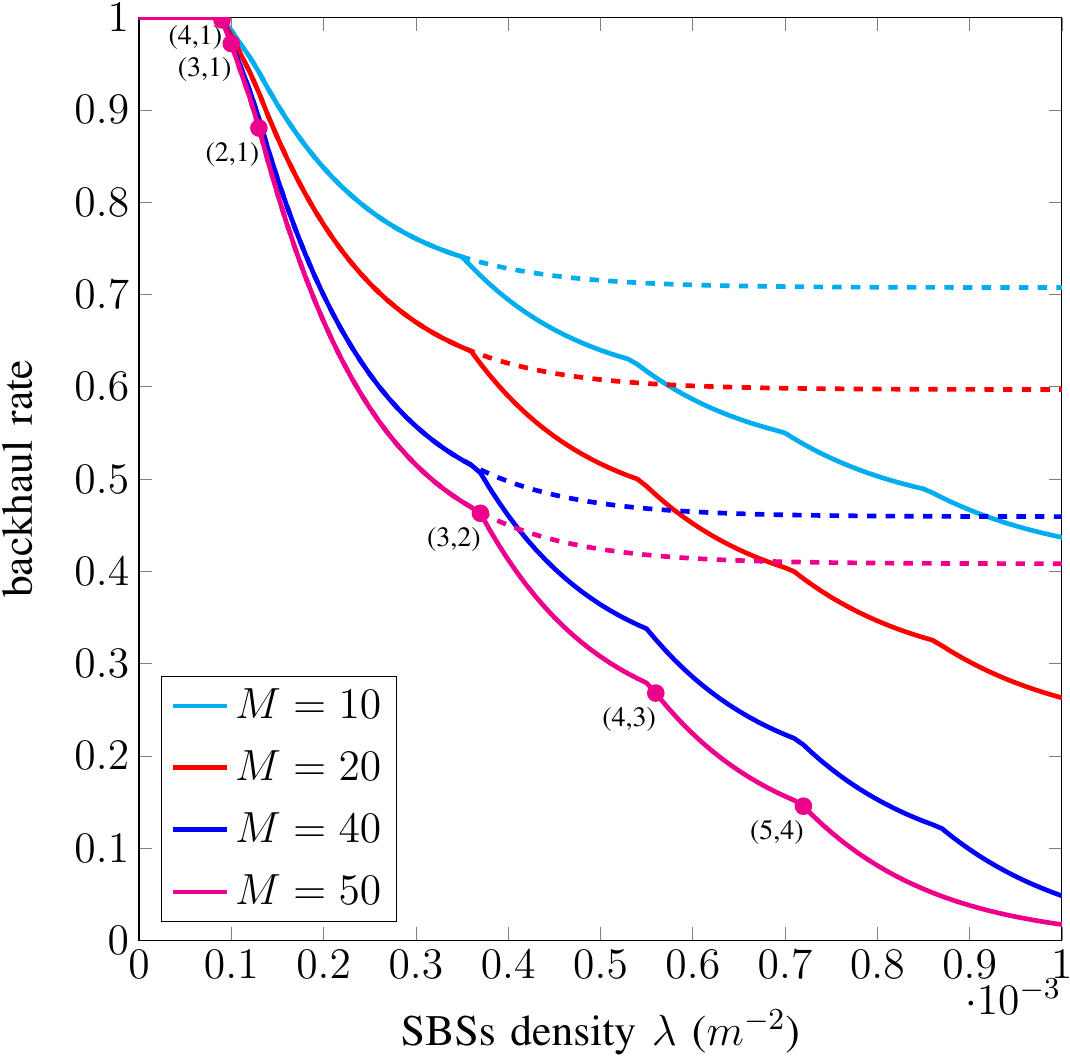}
\caption{Backhaul rate as a function of the density of SBSs $\lambda$ and several values $M$ for the scenario where SBSs are distributed according to a PPP and $T=1$. $F = 200$ files and $\alpha=0.7$. Solid lines correspond to optimal content placement ($\RPIR^*$ in \eqref{eq:optimFixedmuMod}) and dashed lines to popular content placement ($\RPIR^\mathsf{pop}$ in \eqref{eq:popularMod}).}
\label{fig:Density}
\par\end{centering}
\end{figure}

In Fig.~\ref{fig:RWeighted},  we plot the optimized weighted communication rate $\CPIR^*$ in \eqref{eq:optC} for the noncolluding case ($T=1$) as a function of the cache size constraint $M$ and several values of $\theta$. For the considered system parameters, caching is still useful for small values of $\theta$ if the cache size is big enough. For example, for $\theta=0.5$ caching helps in reducing the weighted communication rate with respect to no caching for $M\ge 87$. For $\theta\ge 0.7$, caching does not bring any reduction of the weighted communication rate.
\begin{figure}[!t]
\begin{centering}
\includegraphics[width=\columnwidth]{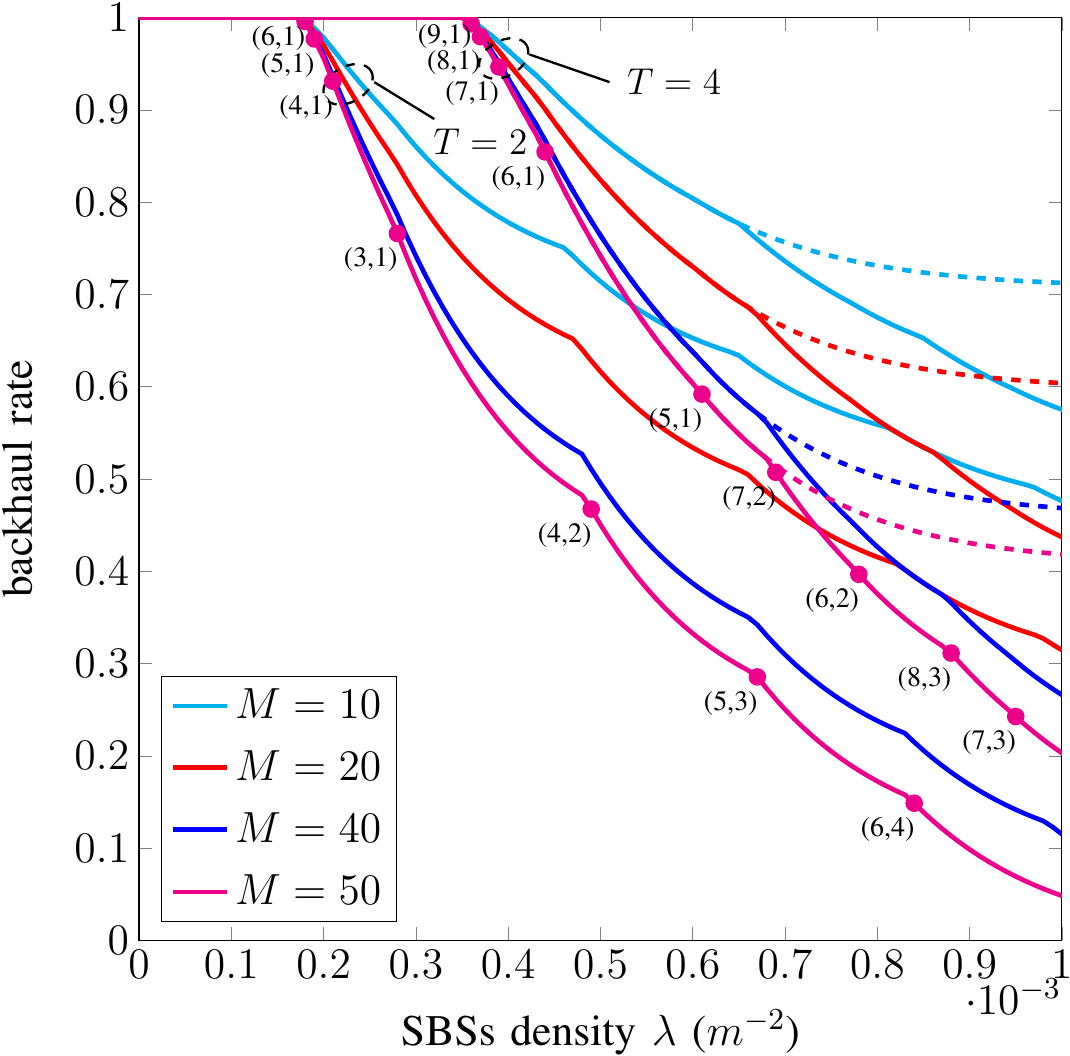}
\caption{Backhaul rate as a function of the density of SBSs $\lambda$ and several values  of $M$ for the scenario where SBSs are distributed according to a PPP and $T=2$ and $T=4$. $F = 200$ files and $\alpha=0.7$. Solid lines correspond to optimal content placement ($\RPIR^*$ in \eqref{eq:optimFixedmuMod}) and dashed lines to popular content placement ($\RPIR^\mathsf{pop}$ in \eqref{eq:popularMod}).}
\label{fig:DensityB}
\par\end{centering}
\end{figure}

In Figs.~\ref{fig:Density} and \ref{fig:DensityB}, we plot the backhaul rate  for a  PPP deployment  model where SBSs are distributed over the plane according to a PPP and a user at an arbitrary location in the plane can connect to all SBSs that are within radius $r_{\mathsf u}$. Let $\lambda$ be the density of SBSs per square meter. For this scenario, the probability that a user is in communication range of $b$ SBSs is given by \cite{Ser17}
\begin{align*} 
\gamma_b = \mathrm{e}^{-\psi}\frac{\psi^b}{b!},
\end{align*}
where $\psi= \lambda \pi {r^2_{\mathsf u}}$. In Fig.~\ref{fig:Density}, we plot the optimized backhaul rate ($\RPIR^*$ in \eqref{eq:optimFixedmuMod}, solid lines) as a function of the density $\lambda$ for $F=200$ files, $\alpha=0.7$, $r_{\mathsf u}=60$ meters, different cache size constraint $M$, and a single spy SBS, i.e., $T=1$. For small densities, caching does not help in reducing the backhaul rate. However, as expected, the required backhaul rate diminishes by increasing the density of SBSs. For comparison purposes, we also plot the backhaul rate for popular content placement ($\RPIR^\mathsf{pop}$ in \eqref{eq:popularMod}, dashed lines). Interestingly, popular content placement is optimal up to a given density of SBSs, after which optimizing the content placement brings a significant reduction of the required backhaul rate. Similar results are observed for $T=2$ and $T=4$ colluding SBSs in Fig.~\ref{fig:DensityB} with the same system parameters as in Fig.~\ref{fig:Density}. In Figs.~\ref{fig:Density} and~\ref{fig:DensityB}, for each $M$ the optimal value of $n$ and $\mu$ depends on the density of SBSs. Typically, a pair $(n,\mu)$ is optimal for a range of densities. In the figures,  we give the optimal values of $n$ and $k$ for $M=50$ (in particular we give the pair $(n,k)$, with $k=1/\nu$, which is also the code parameters of the punctured code $\C'$). For convenience, in the figures we only give the parameters for the densities where the optimal pair $(n,k)$ changes. The values should be read as follows: In Fig.~\ref{fig:Density}, walking the curve from top-left to bottom-right, no caching is optimal for densities up to $\lambda=8\cdot10^{-5}$. For $\lambda=9\cdot10^{-5}$, $(4,1)$ is optimal. Then, $(3,1)$ is optimal for densities $\lambda=10^{-4}$ to $\lambda= 1.2\cdot 10^{-4}$. From $\lambda= 1.3\cdot 10^{-4}$ to $\lambda= 3.2\cdot 10^{-4}$ the optimal value is $(2,1)$, and so on (the curves are plotted with steps of $10^{-5}$).

\section{Conclusion}

We proposed a private information retrieval scheme that allows to download files of different popularities from a cellular network, where to reduce the backhaul usage content is cached at the wireless edge in SBSs, while achieving privacy against a number of spy SBSs. We derived the backhaul rate for this scheme and formulated the content placement optimization. We showed that, as for the no PIR case, up to a number of spy SBSs caching helps in reducing the backhaul rate. Interestingly, contrary to the no PIR case, uniform content placement is optimal. Furthermore, popular content placement is optimal for some scenarios. Although uniform content placement is optimal, the proposed PIR scheme for multiple code rates may be useful in other scenarios, e.g., for distributed storage where data is stored using codes of different rates.

\section*{Appendix\\Proof of \cref{th:CachingScheme}}

To prove that the protocol achieves PIR against $T$ colluding SBSs, we need to prove that both the privacy condition in \eqref{Def: cond1} and the recovery condition in \eqref{Def: cond2} are satisfied. We first prove that the recovery condition in \eqref{Def: cond2} is satisfied. 

According to \cref{lem:GRS}, GRS codes with a fixed weighting vector $\bm v$ and evaluation vector $\boldsymbol{\kappa}$ are naturally nested. Furthermore, puncturing a GRS code results in another GRS code, since GRS codes are weighted evaluation codes \cite[Ch.~5]{HuffmanPless10_1}. Thus, $\C_i' \subseteq \Cmax'$ for all $i$, and it follows from \cref{Eq: Ctilde} that
\begin{align*}
\tilde{\mathcal C} = \left( \sum_{i=1}^F \mathcal C'_i \right) \circ \bar{\mathcal C} =\Cmax' \circ \bar{\mathcal C}.
\end{align*}
Furthermore, it can easily be shown that the Hadamard product of two GRS codes with the same evaluation vector  $(\kappa_1,\ldots,\kappa_n)$ is also a GRS code  with dimension equal to the sum of the dimensions minus $1$. Thus, $\tilde{\mathcal C}$ is a GRS code of dimension $\kmax+T-1$. As $\tilde{\mathcal C}$ is an $(n,\kmax+T-1)$ MDS code (GRS codes are MDS codes), it can correct arbitrary erasure patterns of up to $\Gamma=n-(k_{\mathsf{max}}+T-1)$ erasures. This implies that one can construct a valid $\kmax \times n$ ($d=\kmax$) matrix $\hat{\bm E}$ (satisfying conditions $\mathsf{C1}$--$\mathsf{C3}$) from $\beta=\Gamma$ information sets $\{\mathcal I_m\}$ of $\Cmax'$ as shown below. 
%
%
%

Let $\mathcal{J}_j =  \{j,\ldots,(j+\Gamma-1) \bmod n\}$, $j=1,\ldots,\kmax$. Construct $\hat{\bm{E}}$ in such a way that $\mathcal{J}_j$ is the support of the $j$-th row of $\hat{\bm{E}}$. Hence, $\mathsf{C1}$ is satisfied. Furthermore, since $\tilde{\C}$ is an $(n,\kmax+T-1)$ MDS code and $\Gamma = n-(\kmax+T-1)$, all rows of $\hat{\bm{E}}$ are correctable by $\tilde{\C}$, and thus  $\mathsf{C2}$ is satisfied. 
Finally, run \cref{alg:Ebar}, which constructs $\beta = \Gamma$ information sets  $\{\mathcal I_m\}$ of $\Cmax'$ (and the corresponding sets $\{\mathcal F_l\}$) such that  $\mathsf{C3}$ is satisfied. Note that since $\Cmax'$ is an MDS code, all coordinate sets of size $\kmax$ are information sets of $\Cmax'$, and hence \cref{alg:Ebar} will always succeed in constructing a valid set of information sets of $\Cmax'$ (the inequalities in \cref{line:1,line:2} together with the fact that the overall weight of $\hat{\bm{E}}$ is $\Gamma \kmax$ ensure that $\beta = \Gamma$ valid information sets for $\Cmax'$ are constructed). In particular, the while-loop in \cref{line:1} will always terminate.

From the constructed matrix $\hat{\boldsymbol{E}}$, the user is able to recover  $\Gamma d\ge \beta k_i$ unique code symbols of the requested file
  $\bm X^{(i)}$, at least $k_i$ symbols from each stripe. Furthermore, a set of $k_i$ recovered code symbols from each stripe corresponds to an information set of $\C_i'$ (any subset of size $k_i$ of any information set of size $\kmax$ of $\Cmax'$ is an information set of $\C_i'$), and the requested file $\bm X^{(i)}$ can be recovered. This can be seen following a similar argument as in the proof of \cite[Th.~6]{Kum18}, and it follows that the recovery condition in \eqref{Def: cond2} is satisfied. 
 
Secondly, we consider the privacy condition  in \eqref{Def: cond1}. 
A reasoning similar to the proof of \cite[Lem.~6]{Kum18} shows that it is satisfied, and we refer the interested reader to this proof for further details. The fundamental reason is that addition of a deterministic vector in \eqref{Eq: CollQuery_design} does not change the joint probability distribution of $\{\bm Q^{(l)}, l \in \mathcal{T}  \}$ for any set $\mathcal{T}$ size $T$, and the proof follows the same lines as the proof of \cite[Th.~8]{Hol17b}.  However, note that there is a subtle difference in the sense that independent instances of the protocol may query different sets of SBSs. However, since the set of SBSs that are queried is independent of the requested file and depends only on which SBSs that are within communication range, this fact does not leak any additional  information on which file is requested by the user.

\begin{algorithm}[t]
\SetKwInOut{Input}{Input}
\SetKwInOut{Output}{Output}
\SetKwComment{Comment}{$\triangleright$\ }{}
\DontPrintSemicolon

\Input{$\hat{\bm E}$, $\beta$, $n$, $\kmax$}
\Output{$\{\mathcal{I}_m\}$, $\{\mathcal{F}_l\}$}
%
\For{$m \in \{1,\ldots,\beta\}$}{
$\mathcal{I}_m \leftarrow \emptyset$
}
\For{$l \in \{1,\ldots,n\}$}{
                $\mathcal{F}_l \leftarrow \emptyset$, 
                $m \leftarrow 1$
	        
	        \While{$|\mathcal{F}_l| \leq \Hwt{\bm{t}_l}$}{\label{line:1}
	        \If{ $|\mathcal{I}_m] < \kmax$}{\label{line:2}
	        $\mathcal{F}_l \leftarrow \mathcal{F}_l \cup \{  m\}$\\
	        $\mathcal{I}_m \leftarrow \mathcal{I}_m \cup \{l\}$}
	        
	        $m \leftarrow m+1$
	        
	        }
	        }
	        
	\caption{Construction of $\{\mathcal{I}_m\}$ for Theorem~\ref{th:CachingScheme}  \label{alg:Ebar}}
	
\end{algorithm}

\balance

\bibliographystyle{IEEEtran}
{\small \itemsep 10ex
\bibliography{Bib_PIR.bib}
}

\end{document}